Homo-loggatus. The anthropological condition of historians in the digital world.

Salvatore Spina
salvatore.spina@unict.it

**Abstract**
Computerization has created a digital ecological niche where humans live in a state of interconnection that modifies their Epigenetics. Within this hyper-datafied virtual space, the logged-in agent enhances their intellectual and rational abilities, giving rise to a new cognitive entity. Humans are evolving towards a new anthropological status that shifts the terms of the Digital History debate from History to the historian, compelling the latter to reflect on the positions of Fichte and Schelling regarding the mind-body-world relationship (ecological niche). This reflection leads to the possibility of overcoming the crisis of History imposed by presentism and the necessity of redefining the research methodology based on the new vision of the interconnection between the mind and the digital niche as an investigative tool.

**Reflections through time: humans, machines, Epigenetics.**
The Industrial Revolution begins with a consideration on the reciprocal pact between humans and the environment, which is the foundation of Sciences understood as systems that have marked the fluctuations —if we were to sketch a hypothetical Gaussian curve— of human technological time and scientific thought in enhancement. Machines, a constant product of this reasoning, are being increasingly updated, and their consolidation incessantly modifies our social system, the economic system (thanks to innovation in the production system), the trade system, markets, but also —and above all— the lifestyle of every individual who comes into contact with them.
Every invention is the result of a theoretical, scientific, intellectual path originating in the genius of human beings as thinking beings, who create increasingly innovative tools to facilitate their work and that of others. The subsequent path of the economy is a reflection of this technological enhancement, as is the path of politics (both secular and Christian), which moves from intransigence to censorship to the opportunities offered by engineering products —which become an essential step on which to build national policies of industrial development and innovation.
Every technological achievement is aimed at tempering physical exertion, lifting the body —gradually reducing its power over time— from functions and tasks that can be better performed by a machine. This assertion is a fundamental principle of techno-scientific research, which gave birth to those influences that have been the focus of philosophical "tragedies", which have served little purpose in terms of the danger of constructing a world devoted to the relationship between humans and machines. Marxist concepts such as alienation and loss of one's sense of being have been crucial in tracing the median of a new class of workers who were emerging into History, but who ultimately failed to come to terms with what they despised, subsequently becoming its staunch guardians.
The century of machines, from up-close, becomes a mirror —for itself and the historian— of the millennial path of ingenuity and experimentation, translating into a position —founded on the philosophical courage of Galileo (Villari 1868)— that has destabilized systemic frameworks. Philosophy of Nature frees itself from Humanism and embarks on its own path, free from any ideological presence, accepting only the challenge of an economy that has been looking, from that moment on at technology as the fulcrum of an industrial system whose interest lies in implementing a social and community structure that must provide both labour and a market for the products that workers —and later consumers— produce.
And although Natural Sciences stop where «it is no longer possible to verify or prove» (Giarrizzo 2018), theoretical work and experimentation remain inexhaustible, always followed by practical implementation, necessarily in a machine.
In this sudden rupture, where the divide is swift, History anchors itself to the «inevitable mutations» (Villari 1866), which remain its most important subject, while many branches of Humanities follow the influences of Darwinian theories (Gutiérrez and Ayala 2004; Darwin 2016) and positivist approaches, with an additional push from Mendelian discoveries on the inheritance of traits in living beings, later developed for phenotypic evidence in physical appearance and behavior.
Humans genetically communicate with their territory, with the places from which they derive sustenance and to which they return the surplus. This reciprocal pact implies a reshaping of the phenotypic characteristics of living beings, on which Epigenetics has established its statute: every action has a reaction, which becomes behavioral and physiognomic instruction.
Over the course of millennia, technological advancement has modified our bodies, for example, weakening the muscular and dental structures —responsible for the physical-mechanical relationship with the surrounding space— but on the other hand, there is also evidence of the enhancement of the cerebral system, reaching a structural stage that forms the foundation of our intellectual and rational capabilities, which identify us as the species capable of probing existence and its rules. Science is the manifestation of our increasingly analytical decryption mechanism of the laws of physics, in an attempt to seize the possibilities of changing our surroundings.
On another level, humans have sought to identify their "own" code, in a path that, certainly through Democritus, has found its greatest and definitive explanation in Gregor Mendel (Bateson 2013); and, furthermore, with the discovery of

DNA (Watson and Crick 1953). Since then, scientists have had the opportunity to demonstrate that every single human action unfolds in a genomic-genealogical path that includes the territory, the environment, the community, and ancestors. We are the product of the thinking that has built the tools that have been measuring the world. We are the product of the results achieved, which have transformed manual and subsequently industrial production systems and the economy.

Since the origins of civilization, humans have had to develop technologies and theories aimed at "counting." Disciplines such as economics, demography, anthropology, and sociology are subject to these inventions and become expressions of the attempts by Enlightenment thinkers and humanists to mathematize human action — the only 'object 'that eluded Natural Sciences. And so, just as genetics originated from the work of biologists, a new subfield of mathematics called statistics emerged, aimed at explaining human complexity. However, this investigation required an 'adaptation 'of the instruments that had been available until then to state officials, administrators, and bureaucrats: all Sciences were called to the first tests of a path of "automated computation," in which Pascal and Leibniz played a crucial role. This was the moment when instruments compelled mathematicians, philosophers, and scientists to view society as a new object of research. Malthus engaged in his research by looking at populations, Ricardo at the principles governing income distribution, Morgan at kinship systems, and LePlay at family structures. Everything can be codified and assimilated into second-order abstractions; everything is potentially a number.

**Digital development and Techno-biocenosis.**
Digitalization, while being the result of a convergence of various disciplines, has significantly transformed the ecological niche that has always hosted humans through a continuous process of epigenetic reshaping, starting from their capacity to create communication systems. Human beings encode to communicate. They encode in graphemes to communicate with those who are part of their community; they encode using the same graphemes, but in different combinations, to communicate with their peers who use languages different from their mother tongue; they encode in binary to create discrete formulations to submit to computers, in order to initiate computational processes that enable them to extract information and patterns from data. Today, hyper-informatization and hyper-codification have led humans to develop and utilize a linguistic system that, on one hand, requires specialized training to "interact with the machine," and on the other hand, has made Informatics the syntactic and lexical bridge between various domains of knowledge, both in the physical-natural and humanistic realms, thereby replacing Mathematics, which, after surpassing the interpretative limits imposed by religion, established itself as the language of Natural Sciences. Informatics, now assuming this role, enables interdisciplinary communication between scientific fields, providing Humanities with an "epistemological laboratory" where words and expressions can be analyzed as numbers and mathematical functions, thanks to data mining, machine learning, and other tools. This facilitates the objective structuring of narratives about the past, as seen in the historical sector (Spina 2022). "Digitalization," "Computer Science," "Digitization," "Artificial Intelligence," and "Algorithms" are just a few of the terms that have marked the past decades, during which our physicality has once again demonstrated its role as both the cause and subsequent effect of global changes.

The systems of knowledge and intelligence are transforming. Tools and technologies, once seen as the product of demonic fantasy (Carducci 1964), have become the foundation of every scientific dogma. Simultaneously, the discourse of Epigenetics forces humanists to grapple with the progressive growth of thought models that necessarily link humans to the new computerized reality. Binary encoding materializes humans on a digital dimension, which is an extreme manifestation of their knowledge and turns the world into an expression of the extension of their existence as beings. Everything is given and turned into data. Hyperconnectivity is interconnectivity. That which exists is manifested in the being, a node in the complex communication network; extreme, discrete, syntactic, neural, primary, regardless of the current stage of the symbiotic relationship between humans and the world. Today, we are entering a new phase of our evolution in which our mental faculties, without modifying our bodies, enhance themselves in an external cognitive system beyond our corporeality —which is already the sum of body and mind— but in symbiosis with it. When we had descended from trees millions of years ago, we had begun to encode functions and actions in our phenotype and genetic heritage, just like all living beings on Earth. A lion cub will hunt in the same way as its peers, regardless of the "demonstration" that an adult individual may provide. It is their genetic structure, their cells, their chromosomes, their connections, and physiological structures —their bios— that initiate their functions and drive them to hunt as their species has always done. That bios also pertains to humans. But these initial instructions have been enhanced over time. Additional functionalities have emerged, leading to chromatin modifications (Tsukiyama and Wu 1997; Aalfs and Kingston 2000; Lorch, Maier-Davis, and Kornberg 2010) that allow each "individual" to adapt to events and the passage of time. The bios have been enriched with semantics —the ability to assign meaning to the world**,** and its parts, and to identify the causes of events, regardless of whether a god or a "physical mechanism" is responsible for a lightning strike. What matters is ascribing meaning and seeking the true cause.

Knowledge has become increasingly logical, scientific, and profound, and the Homo-Loggatus (the logged-in human) has deeply apprehended this as the ultimate upgrade in the development of the cognitive subject, a manifestation of the Epigenetics of the Techno-biocenosis, the resulting purpose of digitality —the state of equilibrium and coexistence between humans and technologies.

Digitalization is not merely the translation of an analog signal into a digital one (this goes without saying as a simplistic definition); it is the deconstruction of reality and its migration into an "environment" founded on connectivity. The logged-in human is a subject who lives out their social functions within a structure accessed through specific recognition, a translation of their physical identity into an avatar that retains only informational elements. Being logged in is the necessary status for the expression of their functions. This condition assigns an additional purpose to the body, which combines the external and internal while diminishing its power. Inaction remains analog, while cognition arises solely from the mind and the world. And the world, on its part, is a conjunction of analog and digital information, precisely because it is constantly being adapted to the technologies birthed by human ingenuity (Floridi 2016). It consists of data generated in a dimension whose corporeality is structured by a code composed of elements no longer chemical or physical (carbon, DNA, chromosomes) but rather binary, seeking to emulate the structure of every living being to virtually reproduce the same "subject" without its distinctive physical heritage, the reconstruction and production of which solely remain entrusted to biological and bioinformatic experimentation. What humans restructure in the digital realm, however, is something far more complex: their minds.

To achieve this, the "Homo Technologicus" (the human who creates and uses technology) (Longo 2001; De Toni and Battistella n.d.) modifies the world, adapts it to their inventions, configures it based on the possibilities of machines, and creates all the access keys (such as mobile networks and smartphones) to connect with all of this. All (analog) knowledge converges toward this dimension, which, in an epigenetic sense, has altered "us" and our perception of the triad of body-mind-world, restructuring our psychophysical heritage. It no longer necessitates analysis and understanding solely in an analog and close reading mode but draws directly from repositories that more effectively control all the information we need.

Man adapts to the ecological niche, and his actions elicit reactions from the environment that modify his phenotype, turning that interaction into a sign that can be crucial in the DNA structure. The more radical, profound, and repetitive it is, the stronger the trace becomes, evolving into a physiological and intellectual history, but above all, a new categorical model for the interpretation of inter-relational space (Keverne and Curley 2008; Lester et al. 2011; Roth 2013; Nelson and Monteggia 2011; Crews 2011; Jablonka 2016; Jablonka and Lamb 2002; Dias et al. 2015; Masterpasqua 2009).

The digital era is the new niche where humans reside, a computerized habitat capable of surpassing the concept of "virtual" as a non-specific space of entities, as previously proposed by Lévy (Lévy 1997). Actually, it configures itself as a determined place in both time and space, through coordinates among servers that generate the environment —an even more complex and technologized place, ensuring the equilibrium aimed at the coexistence between humans and machines, between the "existing" human in perpetual intellectual relation with the created-yet-existing-but-without-being device. Hence, all knowledge emerges from this biocenosis, which provides an evolutionary space for a cognitive entity (the Homo-Loggatus) that, thanks to the application of computer tools and Artificial Intelligence, digitally analyzes with greater precision and clarity of detail to infer analogical meanings. The intellectual process lies within this biocenosis and involves every element of the niche, animate and inanimate, mechanical and biological, not necessarily in the Fichtean key concept of the rational self as the sole entity in the cognitive relationship. Every invention has modified space, but digitalization manages to go even further, in a whirlwind of theorizations that, on one hand, make humans the focal point of an information exchange system, while on the other hand, seem to reduce their bodies to its instrumental function —Fichtean (Cogliandro 2011)— in pursuit of the goals that reason sets for itself to achieve, a cornerstone of the development of brain-computer interfaces (Nuyujukian et al. 2018; Willett et al. 2021; Henderson et al. 2005), which allow individuals to act in the world despite living with motor deficits (total or partial) through the implantation of chips in the motor cortex and activation of the "concentration muscle" (Morello 2023).

**Homo-Loggatus Historicus**

The digital ecological niche —Luciano Floridi refers to it as the "infosphere" (Floridi 2020)— is a techno-biological vital space that, on the one hand, binds the cognitive subject to an upgrade, and on the other hand, compels traditional research methodologies to embark on a path that tends towards the re-modulation of disciplinary statutes. While it is true that this invitation has been embraced in many respects —hence, scholars from various humanities disciplines now speak of Digital Humanities— it is even truer that proponents of the digital turn are seeking to impose the logic of individual research domains on processes of digitalization and programming, with an approach that, despite being based on the use of computer machines, remains fundamentally analog, failing to fully grasp the founding principle of digitalization, namely, the encoding into machine-readable language.

In reality, the reflection calls for a different perspective; it necessitates overcoming the digital divide and achieving a dialectical synthesis of the analogue/digital dualism from a linguistic-computerized standpoint: the theory on the infosphere brings us back to the need, as Fogel and Elton already indicated in 1983, to transform "the working [humanist] into a thinking [humanist]" (Fogel and Elton 1983).

If it is true that the living being known as man is entirely thinking, and the parts of his body serve the purpose of acquiring data —beyond mere functional mechanics— in order to process and, thus, carry out formal operations, it is even truer that the Techno-biocenosis is shaping — in an epigenetic sense — a human being who produces digital knowledge in a digital environment with digital tools — which relegate the body to a purely mechanical function — allowing him to organize information into datasets and enabling him to think and signify analogically, but with much deeper and enhanced analytical functions.

Therefore, digitalization is not solely based on the translation from analogue to digital. It is, in fact, the emergence of an ecological niche that surpasses such a dichotomy by shifting the theory from Science to the scholar, from the Humanities to the humanist.

Digital History captures this perspective by shifting the discourse from "History" to the "Historian," who, in an attempt to distance themselves from this perspective, opposes the "present" with a heartfelt and tenacious critique —as articulated by Adriano Prosperi— accusing it of destroying History (Prosperi 2021); meanwhile, François Hartog appeals to "presentism" to describe the process of mesmerization resulting from the profound crisis that, starting with economy, has engulfed humankind manifestations, placing it in a dimension where it no longer recognizes itself because it is incapable of connecting to History and the identities that preceded it; it is a human without origins (Hartog 2015).

The judgment is astute but overlooks the necessity, attempting to deflect it from the redefinition of the "historian's craft" into a digital materialization. It is not presentism, therefore, but the past inability to translate into the narrative of the era of Techno-biocenosis, and the eventfulness that destroys History, doing so through a re-modulation of cognitive entities that demands a progression that historians refuse to undertake.

Therefore, historians feel the entanglement of the web of time in the present, experiencing the loss of reality that increasingly distances itself from the past to connect with the temporal state of the eternal present.

Historians act in the present, narrating what has happened, which is the sum of every action, every moment, every here and now that transforms man into an agent in the impending past, whose action, even though intended for the present, is already historical in its act.

The problem is that one of the effects of hyper-informatization is the drastic and perceivable reduction of the time that elapses between action and its narration as a historical event. This effect produces in historians the illusion of anchoring every event to the present, which almost seems to dissolve the boundary between the present and the past. Acts (understood as something done) no longer require a long period between their occurrence, their passage into the "archive" and their emergence as a useful source for understanding a historical problem. Digitization has reduced the gaps between the event and its entry into History, annihilating the analogue possibilities of historians, who, in the criticism they bring to the ever-present reality, exclude the only element truly in crisis, which is themselves. Historical research is based on traces that must reemerge. Digitization exponentially increases sedimentation, which is the evidence of the acceleration experienced by (human) acts in their ascent to historical sources. Every action is already in the past in its activities, and in the digital ecological niche, the past is rapidly remote and distant, yet it remains in a truly close time. Each action is quickly surpassed, and seriality becomes increasingly complex. Yesterday's sedimentation is already profoundly distant, but it is not obsolete. It remains, even though already in its historical phase, always "present," as the foundational substrate of a new here and now.

Furthermore, digitization, by constantly and rapidly creating for the past in the dynamism of the present, produces inheritable modifications that are leading to cognitive-behavioral changes, whereby new generations will assume an intellectual mechanism that is partly different, not based on the analogue/digital dualism, but on hyper-connection with the Turing Machine.

The digital ecological niche, therefore, is explained by the new anthropological essence of being logged in. And if the "field of historians is the discovery and recording of what actually happened" (Adams 1909), then they necessarily become an Homo-Loggatus, responding to the wonder of digitization, disconnecting the intelligent mind from action, the mind from the mechanics of the body, the ability to act successfully from being intelligent —thus entailing, for them as well, the displacement of their intellectual and formal activities within the digital niche. Here, historical narration expands the concept of sources to include computers, software, and algorithms, which, like every document and monument, are also annotated artifacts, preserved on different media than traditional paper support, thus becoming a testimony of human activity and the humans who created them. However, these "sources" become the unique means to control —and be controlled by!— the enormity of information that arises from the digital environment, while the principle remains unchanged that only the result of human action (the act) becomes a historical source.

The discourse on History, therefore, must shift towards historians and the possibilities of objectivity in their discourse. Complexity and mathematization have deeply influenced History, bringing it closer to Natural Sciences (Holt 1940) with composite and contradictory lines, challenging the established concepts and the role that historians and their methodology had played. Objectivity could —and should— be anchored, as theorized by Leopold von Ranke, to the objective criticism of primary sources and subjecting them to intense epistemological analysis, using techniques from classical Philology (Iggers 1962; Krieger 1977), consequently rejecting the conceptions of History emanating from Moral Philosophy.

Historians had to change their methodology, acquiring different tools and skills. They changed their essence while still attempting to emphasize the unique and particular nature of History and rejecting the applicability of the scientific method to their field of study, which does not concern chemical or physical experiences. This is the path described by Charles H. Hull, who emphasizes the immateriality of the object, asserting that "the ultimate units with which historians deal are not atoms or any type of instrumental abstraction whose individual differences can be ignored, but rather men and the actions of men" (Hull 1914). However, the digital turn imposes a different reflection: historians must grapple with primary units that involve them; the digital niche, the network and interconnections, and computational procedures compel them to become a narrated part of their science, a synthesis of their self-transcendence and aggregation into a new cognitive entity that no longer encompasses the entire analogue complexity that shaped the historians of the previous century. The Homo-Loggatus Historicus (the human logged in history) is precisely the synthesis of such past that must shape the humanists of the new millennium, who face a hurdle to overcome their analogue existence, which must be based on the necessity of a methodology whose depth requires an approach they cannot provide without the assistance of Artificial Intelligence. In narrating the new biocenosis, historians must be part of it, logging their expertise as describers of the present.

Their narration thus becomes strongly logical, anchored to the scientificity of the statement, which does not require sensationalistic rhetoric —for example, speaking of Covid as a "new divine scourge" does not satisfy the historical need for the seriality of the Annales School, nor does it align with the meanings of scientific and datafied argumentation—, capable of dissipating the invisibility (Herbst 1972; Dilthey 1883; Croce 2002) of the non-mechanical elements of History to guide the scholar towards a cognitive existence that de-potentiates the imaginative function (*Ideengeschichte*) that distinguished them. According to Droysen (Droysen 1868; Ries 2010), this function served to connect visible facts with invisible ideas, linking imagination to the fidelity of reality. The digital ecological niche is the visible manifestation of human action, which is not objectified but reveals its ideas and ideologies in behaviors that become perfectly observable, where even illogicality is compressed into definitions and data that explain it. The Homo-Loggatus is a cognitive unit, notably because they relate to a complex and hyper-datafied niche —even though purely physical-quantitative inquiry will not lead to an explanation of human identity— and it can explain its historicity, that is its presence in History *i.e.* the sedimentation of data (analogue and digital) produced by humans; thus, quantification, namely subjectivity is translated into objectivity.

And if the digital niche and the Homo-Loggatus communicate in binary, computational code, a significant turning point is even more necessary for historical discourse, where concepts tend to and must assume the characteristics of formalized and computable statements which are discrete and standardized —a communicative system that, precisely because it aspires to be objective, could overcome the obstacles of historical consciousness as an unsolvable mystery.

What we are today is surpassable. Anthropological evolution is inherent in human nature, and its biocentrism (Schelling 2012; 2018; Sisto 2017) makes it an integral part of the environmental system, which determines its adaptation and epigenetic characteristics (Goldberg, Allis, and Bernstein 2007; Jablonka and Lamb 2014; Weinhold 2006). What we are today is the new phase of evolution, on which Pepperell has reflected (Pepperell 2003), leading humans to an interaction that does not aim to decline them but adapt them —if not now, then in the generation that will follow— to a life that no longer moves solely within analog meanings (by analogy, therefore) but within a network where every individual "Homo" is a semantic node with machines and AI, to which we provide —even unknowingly— data that improves our adaptation to the niche.

**An invitation to reflection**

It is not easy to write the final section of a text that aims to open a debate on multiple fronts.

Digitalization goes beyond historiographic theories but it is in continuity with them. However, in our times, where every single day reminds us that the digital world is a fundamental aspect of the "new" existence, historians —fearing the empowerment of computers and the loss of their control— prefer to comply, fueling a self-referential community, with criticisms of the present and "presentism," rather than becoming the constructive and semantic element of the new cognitive system.

Choosing and attributing meaning will not be the prerogative of AI, algorithms, and other computational tools; "meaning-making" is a human act. The power of intuition derives, indeed, from the possibility of giving meaning to the world, but it rests on the principle of arbitrariness —though not illogicality— which does not belong to the Turing Machine, as it is an intellectual opportunity that does not arise from simple computation. This function belongs to humans and, even more so, to historians.

However, humanists seem to underestimate the opportunity for a dialogue that can establish a connection between History and Information technologies. And if this was true sixty years ago —when strong skepticism turned into evident pessimism (Judt 1979), in an attempt not to lose political control of the field of historical research— it is even more true today.

"No meet this bar," Milligan wrote in 2019 (Milligan 2019), highlighting the difficulty and closure of a generation of historians who looked at their theoretical positions but without questioning —the absence of innovation— and the status of History.

From the historiographic theories of the 19th century, which were at the core of Historical Science —from Wilhelm von Humboldt (Humboldt 1990) to Leopold von Ranke (Ranke and Ramonat 2010) and Marc Bloch (Bloch 1963)— there has been no innovation in the epistemic approach, which still relies on the concept of "research without a laboratory," all based on the scholar's intuition (Spina 2022). However, the definitions of an epistemology of History are a proof of a constant determination to build a systematic structure for its status, which has allowed historians to connect the Philosophies of Nature, with the epistemic-laboratory sciences.

For this reason, historians cannot sever the dialogue with digital innovation, as it is precisely the latter that has the necessary tools to bring historical knowledge to a state of objectivity that is impossible to achieve without such computational instruments, which, in terms of multidisciplinarity, represent the new linguistic system of Knowledge, allowing History to be an integral part of the scientific debate.

Hence, what has appeared here is an imbalance in ideological positions.

Even today, while speaking of the digital world, historians seek refuge behind their pessimism, trying to escape the digital niche and the emergence of the Techno-biocenosis. If, for "digital" historians, computers and calculators are the best tools to study human actions, complex societies, and the dynamics that have marked significant and minor historical events, for the "analog" historians, they are merely information collectors, replacing typewriters and enabling the visualization of appropriately photographed archival documents.

The development of computational analysis, visualization, and interpretation of linguistic meanings (the binding force of all large and small communities), which allows scholars to view humans as nodes in a network of correlations, does not destabilize the positions of traditionalist historians, who continue to ignore the positions of Le Roy Ladurie and the need for proper training that allows historians (Le Roy Ladurie 1968; Rosenberg 2003) to fully understand the Hyperlearning revolution (Perelman 1992) and the Digital Divide (Dougherty and Nawrotzki 2016).

The great task of historians is to build a new *Heimat,* in which the bond between humans and computers becomes systemic (Jouman Hajjar 2021; Pati 2021; Zhavoronkov 2021) to re-ontologize the world and be able to catalyze historical knowledge towards logicality and objectivity.

Historians must understand the purpose of the Turing Machine, its "inner" mechanism, and the need to assemble the knowledge that can be reshaped in its linguistics, meanings, and representations, in order to combine the analog and digital worlds. They must, therefore, play the role of the keystone in the coexistence system of meanings, where "all the discourses [...] that are the principles of time" are hosted (Hegel 2022), at a time when the projection of digitality is directed towards the (near) future, in an space-time continuum where historians will inevitably live in a symbiotic wireless interconnection with the Turing Machine, which will provide Artificial Intelligence tools —think of Transkribus (Muehlberger et al. 2019) and ChatGPT (Zhai 2022; Pavlik 2023; Alshater 2022)— and computer systems that will assist the Homo-Loggatus Historicus in translating data and information into a more objective narrative of the past and in having more effective control over its dissemination, at a time when, amidst polarizations and post-truth politics, History has become an open field, and network communication is entirely based on the ideology of Silicon Valley.